\documentclass[journal,twoside,web]{ieeecolor}
\usepackage{lcsys}

\usepackage[utf8]{inputenc}
\usepackage[T1]{fontenc}
\usepackage[english]{babel}

\usepackage{mathtools}
\usepackage{amssymb} 

\usepackage{bm} 

\usepackage{newtxtext,newtxmath}  

\usepackage[version=3]{mhchem} 
\usepackage{esint} 
\usepackage{dsfont} 
\usepackage[makeroom]{cancel} 
\usepackage{empheq}
\usepackage{physics}
\usepackage[binary-units=true,per-mode=symbol]{siunitx}

\usepackage{graphicx}
\usepackage{float}

\usepackage{subcaption}

\usepackage{booktabs}
\usepackage{makecell}
\usepackage{threeparttable}
\usepackage{multirow}
 
\usepackage[dvipsnames]{xcolor}

\usepackage[nameinlink,capitalise,noabbrev]{cleveref}
\crefformat{equation}{(#2#1#3)}
\Crefformat{equation}{Equation (#2#1#3)}
\crefrangeformat{equation}{(#3#1#4) to~(#5#2#6)}
\crefmultiformat{equation}{(#2#1#3)}%
{ and~(#2#1#3)}{, (#2#1#3)}{ and~(#2#1#3)}
\Crefmultiformat{equation}{Equations (#2#1#3)}%
{ and~(#2#1#3)}{, (#2#1#3)}{ and~(#2#1#3)}

\usepackage[noadjust]{cite} 
\newcommand*{\citet}[1]{\cite{#1}}

\usepackage{pgf,tikz,pgfplots}
\pgfplotsset{compat=1.10}
\usetikzlibrary{shapes,arrows,calc,arrows.meta,
patterns,backgrounds,positioning,fit}
\tikzset{every picture/.style=black}
\usepackage{circuitikz}

\tikzset{three sided/.style={
        draw=none,
        append after command={
            [shorten <= -0.5\pgflinewidth]
            ([shift={(-1.5\pgflinewidth,-0.5\pgflinewidth)}]				  \tikzlastnode.north east)
        edge([shift={( 0.5\pgflinewidth,-0.5\pgflinewidth)}]				  \tikzlastnode.north west) 
            ([shift={( 0.5\pgflinewidth,-0.5\pgflinewidth)}]				  \tikzlastnode.north west)
        edge([shift={( 0.5\pgflinewidth,+0.5\pgflinewidth)}]				  \tikzlastnode.south west)            
            ([shift={( 0.5\pgflinewidth,+0.5\pgflinewidth)}]				  \tikzlastnode.south west)
        edge([shift={(-1.0\pgflinewidth,+0.5\pgflinewidth)}]				  \tikzlastnode.south east)
        }
    }
}

\usepackage{psfrag}
\usepackage{pstool} 

\Crefformat{figure}{#2Fig.~#1#3}
\Crefmultiformat{figure}{Figs.~#2#1#3}{ and~#2#1#3}{, #2#1#3}{ and~#2#1#3}

\makeatletter
\newcommand{\getfontsize}{\f@size pt}
\makeatother

\newcommand*\ds{\displaystyle}

\newcommand*\diff{\mathop{}\!\mathrm{d}}

\newcommand*\VDD{V_{\mathrm{DD}}}

\newcommand*\VL{V_{\mathrm{L}}}
\newcommand*\VH{V_{\mathrm{H}}}
\newcommand*\VM{V_{\mathrm{M}}}

\newcommand*\vOUTi{v_{1}}
\newcommand*\vOUTii{v_{2}}

\renewcommand*\vv{v}

\newcommand*\Dvv{\Delta v}

\newcommand*{\EE}{\mathcal{U}}
\newcommand*\sigmaw{\sigma}

\newcommand*\sigmawo{\sigma_{0}}
\newcommand*\sigmawM{\sigma_{\mathrm{M}}}

\newcommand*{\EEav}{\overline{\EE}}

\newcommand*\tauo{\tau_{0}}
\newcommand*\tauM{\tau_{\mathrm{M}}}

\newcommand*\MTT{MTT}

\newcommand*\hh{\tilde{h}}

\definecolor{k}{rgb}{0 0 0}
\definecolor{r}{rgb}{1 0 0}
\definecolor{g}{rgb}{0 1 0}
\definecolor{b}{rgb}{0 0 1}
\definecolor{orange}{rgb}{1,0.7,0}
\definecolor{c}{rgb}{0 1 1}
\definecolor{cc}{RGB}{64 224 208}
\definecolor{m}{rgb}{1 0 1}
\definecolor{khaki}{RGB}{128 128 0}
\definecolor{deepskyblue}{RGB}{0 191 255}
\definecolor{darkMagenta}{rgb}{0.5 0 0.5}
\definecolor{chocolateBrown}{RGB}{98 52 18}
\definecolor{lightBrown}{RGB}{189 154 122}
\definecolor{mybrown}{RGB}{127 37 0}
\definecolor{bordeaux}{RGB}{131 41 85}
\definecolor{violet}{RGB}{127 0 255}
\definecolor{myGreen}{RGB}{134,180,44}
\definecolor{gray_gate}{RGB}{211,208,205}
\definecolor{yellow_oxide}{RGB}{244,231,164}
\definecolor{color_mix}{rgb}{0.7510 0.2510 0.2510}

\definecolor{h}{rgb}{0 0 0}

\definecolor{l}{rgb}{0 0 1}

\graphicspath{{figures/}}

\def\BibTeX{{\rm B\kern-.05em{\sc i\kern-.025em b}\kern-.08em
    T\kern-.1667em\lower.7ex\hbox{E}\kern-.125emX}}
\begin{document}

\title{Predicting State Transitions in Autonomous Nonlinear Bistable Systems with Hidden Stochasticity}
\author{Léopold Van Brandt and Jean-Charles Delvenne
\thanks{The work has been partially supported by the Research Project "Thermodynamics of Circuits for Computation" of the National Fund for Scientific Research (FNRS) of Belgium.}
\thanks{The authors are with 
UCLouvain, Louvain-la-Neuve, Belgium (e-mail: leopold.vanbrandt@uclouvain.be, jean-charles.delvenne@uclouvain.be).}}


\maketitle

\thispagestyle{empty} 
   
\begin{abstract} 

Bistable
autonomous systems can be found in many areas of science. 
When the intrinsic noise intensity is large, these systems exhibits stochastic transitions from one metastable steady state to another. 
In electronic bistable memories, these transitions are failures, usually simulated in a Monte-Carlo fashion at a high CPU-time price.
Existing 
closed-form
formulas,
relying on near-stable-steady-state approximations of the nonlinear system dynamics to estimate the mean transition time, have turned out inaccurate.
Our contribution is twofold.
From a unidimensional stochastic model of overdamped autonomous systems, we propose an extended Eyring-Kramers 
analytical
formula accounting for both nonlinear drift and state-dependent white noise variance, rigorously derived from Itô stochastic calculus.
We also 
adapt it to practical system engineering 
situations
where the intrinsic noise sources are hidden and can only be 
inferred from the fluctuations of observables measured in 
steady states.
First numerical trials on an industrial 
electronic 
case study suggest that our approximate prediction formula achieve remarkable accuracy, outperforming 
previous 
non-Monte-Carlo approaches.

\end{abstract}



\section{Introduction}

\begin{figure}
\captionsetup[subfigure]{singlelinecheck=off,justification=raggedright} 
\newcommand\myfontsize{\small}
\captionsetup[subfigure]{skip=0pt}
\begin{subfigure}[t]{\linewidth}
\centering
\subcaption{}
\label{fig_latch}
{\footnotesize\vspace{-\baselineskip}}
\begin{circuitikz}[american voltages, transform shape, line cap=rect, nodes={line cap=butt},scale=1]
\newcommand\mycolor{black}
\newcommand\dy{1.5}
\draw
(0,0) node[not port,color=\mycolor,scale=1.0] (inv) {}
++ (0,-\dy) node[not port,scale=-1.0,color=\mycolor] (inv2) {};
\draw[\mycolor]
(inv.out) 
-- ++(+1,0) node[label={[font=\normalsize,color=red]above:
$\vOUTi(t)$
}] (voutt1) {\color{r}$\bullet$}
-- ++(0,-0.5*\dy) node[] (vout1) {}
-- ++(0,-0.5*\dy) node[] (vout1b) {}
-- (inv2.in);
\draw[\mycolor]
(inv2.out) 
-- ++(-1,0) node[] (vout2) {};
\draw[\mycolor]
(inv.in) 
-- ++(-1,0) node[label={[font=\normalsize,color=gray]above:
$\vOUTii(t)$
}] (vin1e) {\color{gray}$\bullet$};
\draw[\mycolor]
(vout2.center) -- (vin1e.center);
\end{circuitikz}
\end{subfigure}
\captionsetup[subfigure]{skip=2pt}
\begin{subfigure}[t]{\linewidth}
\newcommand\colorX{b}
\centering
\myfontsize 
\subcaption{}
\label{fig_MTNS2024_NOISETRAN_VDD_70mV}
\psfragscanon
\psfrag{t [ms]}[cc][cc]{$t \, [\si{\milli\second}]$}
\psfrag{v(t) [mV]}[cc][cc]{$\textcolor{r}{\vOUTi(t)}$, $\textcolor{gray}{\vOUTii(t)}$}
\psfrag{-10}[cc][cc]{$-10$}
\psfrag{0}[cc][cc]{$0$}
\psfrag{1}[cc][cc]{$1$}
\psfrag{2}[cc][cc]{$2$}
\psfrag{3}[cc][cc]{$3$}
\psfrag{4}[cc][cc]{$4$}
\psfrag{5}[cc][cc]{$5$}
\psfrag{6}[cc][cc]{$6$}
\psfrag{8}[cc][cc]{$8$}
\psfrag{10}[cc][cc]{$10$}
\psfrag{20}[cc][cc]{$20$}
\psfrag{30}[cc][cc]{$30$}
\psfrag{40}[cc][cc]{$40$}
\psfrag{50}[cc][cc]{$50$}
\psfrag{60}[cc][cc]{$60$}
\psfrag{70}[cc][cc]{$70$}
\psfrag{80}[cc][cc]{$80$}
\psfrag{vOUT2(t)}[cl][cl]{$\textcolor{gray}{\vOUTii(t)}$}
\psfrag{vOUT1(t)}[cl][cl]{$\textcolor{r}{\vOUTi(t)}$}
\psfrag{VL}[cc][cc]{\color{\colorX} $\VL$}
\psfrag{VM}[cc][cc]{\color{\colorX} $\VM$}
\psfrag{VH}[cc][cc]{\color{\colorX} $\VH$}
\includegraphics[scale=1]{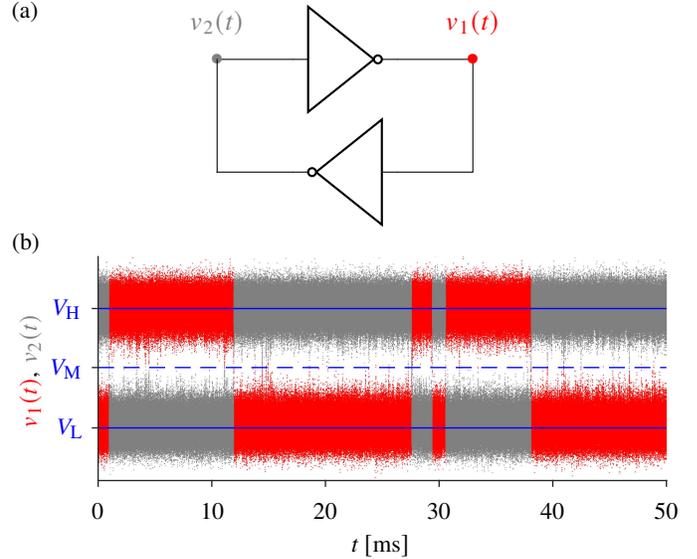}
\end{subfigure}
\caption{
\subref{fig_latch} SRAM bitcell retaining a logical 0 or 1 encoded on two complementary node voltages ($\vOUTii$ and $\vOUTi$) as low and high levels $\VL$ and $\VH$.
The retained state is stabilised by a feedback loop implemented by two cross-coupled inverters.
An inverter is a nonlinear time-invariant system producing a high $\VH$ (resp. low $\VL$) output when its input is maintained at constant low $\VL$ (resp. high $\VH$), yet with internal dynamics and intrinsic noise.
\newline
\subref{fig_MTNS2024_NOISETRAN_VDD_70mV} Transient noise simulation at supply voltage $\VDD = \SI{70}{\milli\volt}$ (adapted from \cite{SSE2023}). Intrinsic noise-induced stochastic state transitions (bit flips  $\VL \leftrightarrow \VH$) are observed. $\VM$ denotes the threshold voltage corresponding to the unstable state.
\textcolor{h}{For the illustrated case, the bistable system is symmetrical in the sense that two inverters are identical, making the two steady states equiprobable and the transitions $\VL \leftrightarrow \VH$ rates equal.}
}
\label{fig_SRAM}
\end{figure}

\IEEEPARstart{B}{istable} 
 autonomous systems are ubiquitous in many scientific fields such as
biology~\cite{Dhilnikov2005,Goldbeter2018}, climatology~\cite{Livina2010,Hirota2011}, chemical reaction systems~\cite{Vellela2009}, and electronics.
Our chief example in this paper is Static Random Access Memory (SRAM) bitcells (\cref{fig_latch}), embedded in manufactured chips for temporary information storage~\cite{SSE2023,LASCAS2024,EDTM2024}.
Generally, bistable systems admit, by definition, two stable 
steady states, sometimes referred to as 
“metastable” points~\cite{Vellela2009,Berglund2011,Freitas2022_reliability} insofar as \emph{stochastic transitions} from one state to the other are observed due to random noise intrinsic to the dynamics~\cite{Rezaei2020,Freitas2022_reliability} (\cref{fig_MTNS2024_NOISETRAN_VDD_70mV}). 

The problem addressed in this paper is to find out the mean time to switch from one stable state to another, driven by random noise.
 This time (defined rigorously later) is the \emph{mean transition time} ($\MTT$).
Characterising the statistics of the transition time, notably the $\MTT$, is indeed crucial for many encountered bistable systems, especially \textcolor{h}{SRAM bitcells in which the state transitions mean loss of 
the originally
retained 
bit
and hence are 
failures~\cite{Rezaei2020,SSE2023,LASCAS2024,EDTM2024}}.
The naive approach is to empirically estimate the $\MTT$ from dynamic Monte-Carlo simulation, whose major weakness is its computational intensity (since statistical relevance requires repeating or lengthening transient experiment like \cref{fig_MTNS2024_NOISETRAN_VDD_70mV}).
As reviewed in \cref{section:Previous Works}, several alternative approaches, all faster but with different accuracy/insight trade-offs, have been proposed in electronics in response to the practical need for efficient prediction of SRAM bitcell reliability.
At the same time, the calculation of the $\MTT$ is a classical problem in chemical physics~\cite{Weidenmuller1984,Berglund2011}, mathematically formalised as a \emph{first-passage-time} (FPT) problem~\cite{Siegert1951,Nobile1985}.
The celebrated \emph{Eyring–Kramers law} was first derived from unidimensional stochastic calculus in~\cite{Kramers1940} and generalised only in the early 2000s (see for instance~\cite{Berglund2011,Bouchet2016}). 
In the present work, we aim at explaining how it can be suitably extended and applied to accurately predict the $\MTT$ in SRAM bitcells, from cheap deterministic (non-Monte-Carlo) circuit simulations. The framework is the stochastic nonlinear model previously developed in~\cite{EDTM2024}.

Importantly, we emphasise that, while working under the convenient white Gaussian noise assumption, the variance of the noise inherent to the dynamics generally depends on the state of the system.
It cannot therefore be merely considered as constant for accurate $\MTT$ predictions. 
Furthermore, this variance may not be explicitly known in an analytical form, or even measurable in
non steady 
states for many physical systems, when the 
different intrinsic noise sources of various origins are partially 
unknown or
hidden.
A simple example is Johnson-Nyquist noise due to the random thermal agitation of the charge carriers in dissipative electronic devices~\cite{Johnson1928,Nyquist1928}.
The celebrated fluctuation–dissipation theorem~\cite{Kubo1966} applies only for a linear resistor, but not for arbitrary nonlinear devices operating out-of-equilibrium~\cite{APL2023}.
More generally, models describing the variations of the noise statistics, notably the variance, with the state are not 
straightforwardly
available for complex nonlinear systems operating far-from equilibrium.
They are sometimes hidden in the industrial model of the system components (e.g. transistors within the inverter blocks of the SRAM bitcell in \cref{fig_latch}).
This is why in this work, we also adapt our predictive developments to the generic situation where the noise intensity can only be extracted, from fluctuations of observables, at steady states of the system.

The rest of this paper is structured as follows.
\Cref{section:Previous Works} further describes the SRAM bitcell, exemplifying bistable systems.
We provide a reminder of the equations governing its nonlinear dynamics (\cref{subsection:SRAM Bitcells Exemplifying Multidimensional Autonomous Nonlinear Bistable Systems}), 
present a general unidimensional stochastic nonlinear model of bistable systems (\cref{subsection:Unidimensional Stochastic Nonlinear Model}),
and review previous attempts of numerical or analytical predictions of the $\MTT$, relying on near-steady-state approximations (\cref{subsection:Near-Stable-Steady-State Approximate Formulas}).
Notably, \cref{subsection:Classical Eyring-Kramers Formula} reintroduces Kramers formula following~\cite{Berglund2011}, which still assumes constant noise intensity.
Our main result is derived and detailed in \cref{section:Stochastic Formulation for Varying Noise Intensity}: from Itô 
stochastic calculus
(\cref{subsection:Itô Drift-Diffusion Process}), we address the more general case of state-dependent noise intensity and provide
an extended explicit formula
to compute the $\MTT$ through deterministic numerical integration (\cref{subsection:Extended Eyring-Kramers Formula}).
We also propose an approximate extended Eyring-Kramers formula involving only the noise variance extracted at the stable and unstable steady states (\cref{subsection:Approximate Extended Eyring-Kramers Formula}).
\Cref{section:Predictions and Discussion} compares and discusses the accuracy of different closed-form formulas applied to SRAM bitcells, under the practical constraint of partially 
observable
stochasticity.
\Cref{section:Conclusions and Perspectives} finally draws the conclusion and opens perspectives.

\section{Previous Works}
\label{section:Previous Works}

\subsection{SRAM Bitcells Exemplifying Multidimensional Autonomous Nonlinear Bistable Systems}
\label{subsection:SRAM Bitcells Exemplifying Multidimensional Autonomous Nonlinear Bistable Systems}

In retention mode, an SRAM bitcell is undriven and isolated from peripheral circuitry in order to retain a bit of data.
It can be assimilated to a cross-coupled inverter pair (\cref{fig_latch}), a bistable electronic circuit implemented with four transistors.

The conventional state-space representation involves two node voltages \textcolor{h}{$\bm{v} = (\vOUTi,\vOUTii)$} as states~\cite{Zhang2006}.
The \emph{stochastic state-space} representation of the SRAM bitcell of \cref{fig_latch} can be written as an 
{\color{h}
Itô's \emph{stochastic differential equation} (SDE):
\begin{equation}
\label{eq:2D}
\bm{C}(\bm{v}) \diff{\bm{v}} = \bm{I}(\bm{v}) \diff{t} + \bm{s}(\bm{v}) \diff{\bm{W}}(t)
\text{.}
\end{equation}
In \eqref{eq:2D}, the vector $\bm{I}(\bm{v}) = (I_1(\vOUTi,\vOUTii),I_2(\vOUTi,\vOUTii))$ denotes the mean currents flowing out of the inverters, a nonlinear function of both state variables $\vOUTi$ and $\vOUTii$.
The matrix $\bm{C}(\bm{v})$ has the dimensions of a capacitance and relates the voltage variations to the currents. 
The vector $ \bm{W}(t) = (W_1(t),W_2(t))$ contains two standard Wiener processes, i.e. the integrals of independent white Gaussian noise processes, and is multiplied by the square matrix $\bm{s}(\bm{v})$, here modelling the thermal noise of the transistors.

Because matrix $\bm{C}$ is known to be invertible (as diagonally dominant) from physical considerations,
we may 
transform \eqref{eq:2D} into
\begin{equation}
\label{eq:dv/dt matrix}
\begin{aligned}
\diff{\bm{v}}
&= \bm{C}(\bm{v})^{-1} \bm{I}(\bm{v}) \diff{t} + \bm{C}(\bm{v})^{-1} \bm{s}(\bm{v}) \diff{\bm{W}}(t) \\
&\equiv 
\bm{h} (\bm{v}) \diff{t} + \bm{\sigma}(\bm{v})  \diff{\bm{W}}(t)
\text{.}
\end{aligned}
\end{equation}
to retrieve the canonical form of state-space representation of nonlinear dynamical systems, if needed.


}

The bistable system has two stable steady states 
($(\VL,\VH)$ and $(\VH,\VL)$ with voltage levels shown in \cref{fig_MTNS2024_NOISETRAN_VDD_70mV}) and one unstable state ($\vOUTi = \vOUTii = \VM$)~\cite{Vellela2009,Hirota2011,Goldbeter2018,LASCAS2024,EDTM2024}.
{\color{h}
For the special case illustrated in \cref{fig_SRAM}, the two inverters are identical, hence the roles of $\vOUTi$ and $\vOUTii$ are interchangeable in \eqref{eq:2D} (e.g. $I_1(\vOUTi,\vOUTii) = I_2(\vOUTii,\vOUTi)$). Consequently, 
the two steady states are equiprobable, and the transitions rates are equal. In that situation, the bistable system is said symmetrical and only one single $\MTT$ must be predicted.
Nevertheless, all the methodology that follows is intended to be perfectly general, also covering the asymmetrical case with two different transition rates.
}

Although the $\MTT$ can be 
estimated from Monte-Carlo simulation on a digital twin (\cref{fig_MTNS2024_NOISETRAN_VDD_70mV}, 
performed on
an industrial circuit simulator), previous works \cite{LASCAS2024,EDTM2024} showed that this approach is prohibitively expensive.
Relying on the formulation \eqref{eq:2D} 
yet with
constant capacitances 
$\bm{C}$,
reference \cite{Rezaei2020} developed an accelerated noise simulator, based on an incremental algorithm with successive thresholds, for fast SRAM reliability assessment.
Its difficult extension to other electronic memory technologies, even more to other bistable systems of different fields is however a major limitation calling for more general non-Monte-Carlo approaches.

\subsection{\color{h}Unidimensional Stochastic Nonlinear Model}
\label{subsection:Unidimensional Stochastic Nonlinear Model}

\begin{figure}[]

\vspace{3mm}

\captionsetup[subfigure]{singlelinecheck=off,justification=raggedright}
\newcommand\myfontsize{\small}
\newcommand\mytickfontsize{\footnotesize}
\captionsetup[subfigure]{skip=0pt}
\begin{subfigure}[t]{\linewidth}
\myfontsize
\centering
\psfragscanon
\psfrag{dvv/dt [mV/us]}[cc][bc]{\myfontsize$h(\vv) \, [\si{\milli\volt\per\micro\second}]$}
\psfrag{-0.05}[cc][cc]{\mytickfontsize$-0.05$}
\psfrag{0}[cc][cc]{\mytickfontsize$0$}
\psfrag{0.05}[cc][cc]{\mytickfontsize$0.05$}
\includegraphics[scale=1]{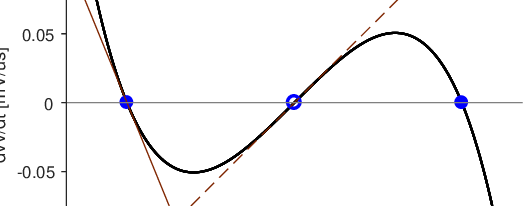}
\vspace{-3.5cm}
\vspace{-\baselineskip}
\subcaption{}
\label{fig_MTNS2024_h}
\end{subfigure}

\vspace{+1mm}

\captionsetup[subfigure]{skip=0pt}
\begin{subfigure}[t]{\linewidth}
\myfontsize
\centering
\psfragscanon
\psfrag{vv [mV]}[cc][cc]{\myfontsize$\textcolor{violet}{\vv} \, [\si{\milli\volt}]$}
\psfrag{E(vv) [V2/s]}[cc][cc]{\myfontsize$\EE(\vv) \, [\si{\square\volt\per\second}]$}
\psfrag{-10}[cc][cc]{\mytickfontsize$-10$}
\psfrag{0}[cc][cc]{\mytickfontsize$0$}
\psfrag{0.5}[cc][cc]{\mytickfontsize$0.5$}
\psfrag{1}[cc][cc]{\mytickfontsize$1$}
\psfrag{1.5}[cc][cc]{\mytickfontsize$1.5$}
\psfrag{10}[cc][cc]{\mytickfontsize$10$}
\psfrag{15}[cc][cc]{\mytickfontsize$15$}
\psfrag{20}[cc][cc]{\mytickfontsize$20$}
\psfrag{25}[cc][cc]{\mytickfontsize$25$}
\psfrag{30}[cc][cc]{\mytickfontsize$30$}
\psfrag{35}[cc][cc]{\mytickfontsize$35$}
\psfrag{40}[cc][cc]{\mytickfontsize$40$}
\psfrag{50}[cc][cc]{\mytickfontsize$50$}
\psfrag{60}[cc][cc]{\mytickfontsize$60$}
\psfrag{70}[cc][cc]{\mytickfontsize$70$}
\psfrag{80}[cc][cc]{\mytickfontsize$80$}
\psfrag{90}[cc][cc]{\mytickfontsize$90$}
\includegraphics[scale=1]{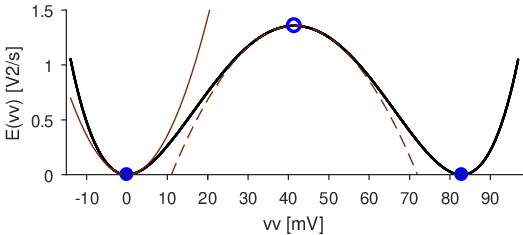}
\vspace{-4cm}
\vspace{-\baselineskip}
\vspace{-1mm}
\subcaption{}
\label{fig_MTNS2024_E}
\end{subfigure}
\caption{
\subref{fig_MTNS2024_h} Deterministic drift term in \eqref{eq:dvv/dt}, deriving from \subref{fig_MTNS2024_E} quasi potential $\EE(\vv) = - \int_{0}^{\vv} h(\vv') \,\diff{\vv'}$.
Near-stable-steady-state (full blue dots) approximations inherent to \cite{Kish2002,Nobile1985}'s formulas are shown in full brown lines.
Eyring-Kramers formula also exploits the behaviour near the unstable point (open blue dot and dashed brown curve).
\newline
Illustrated case: same as \cref{fig_MTNS2024_NOISETRAN_VDD_70mV} (SRAM bitcell at $\VDD = \SI{70}{\milli\volt}$).
Extracted
infinitesimal variance at 
steady states: 
$\sigmawo = \SI{258}{\square\volt\per\second}$ (stable) and $\sigmawM = \SI{105}{\square\volt\per\second}$ (unstable).
}
\label{fig_h_E}
\end{figure}

Upon observation of transient experiences like  \cref{fig_MTNS2024_NOISETRAN_VDD_70mV} and of time-scale separation, we have proposed to reduce the dimensionality of system \eqref{eq:dv/dt matrix} by introducing an unidimensional (1D) reaction coordinate~\cite{LASCAS2024,EDTM2024},
\textcolor{h}{
(e.g. $\vv = (\vOUTii -\vOUTi)/\sqrt{2}$ is adequate for the case illustrated in \cref{fig_SRAM})
}
The 
\textcolor{h}{vocable}
reaction coordinate comes from chemistry~\cite{Rogal2021}.
This is equivalent to studying the nonlinear system dynamics in the main direction dominating the transition time, described by the 
{\color{h}
unidimensional
Itô's SDE
\begin{equation}
\label{eq:dvv/dt}
\diff{\vv}
= h(\vv) \diff{t} + \sigmaw(\vv) \diff{W}(t)
\text{.}
\end{equation}
}
\Cref{eq:dvv/dt}
models
an overdamped single-state system with \emph{drift-diffusion} dynamics, like a Brownian particle~\cite{Berglund2011,Bouchet2016}.

The nonlinear drift function $h(\vv)$ (\cref{fig_MTNS2024_h}) is understood as describing the natural dynamics of the system in absence of noise, 
here cheaply extracted from a deterministic dynamic simulation~\cite{EDTM2024}. It derives from a scalar \emph{quasi potential} $\EE(\vv)$~\cite{Berglund2011,Bouchet2016} (\cref{fig_MTNS2024_E}): $h(\vv) \equiv -\diff{\EE}/\diff{\vv}$.
The stable steady states ($\vv = 0$ and $\vv = 2\Dvv$) correspond to the valleys of $\EE(\vv)$, whereas the 
hill
characterises the unstable point ($\vv = \Dvv$, labelled $\mathrm{M}$). 

In \eqref{eq:dvv/dt}, the diffusion 
is the white Gaussian noise $\sigmaw(\vv) W(t)$: $\sigmaw^2(\vv)$ is the infinitesimal variance, or variance per unit time (i.e. in $\si{\square\volt\per\second}$) and  depends on the state $v$ 
for most nonlinear systems;
\textcolor{h}{$W(t)$ is the Wiener process
(in $\si{\second^{1/2}}$)}.
All previous works reviewed below coarsely assumed that $\sigmaw^2(\vv)$ is \emph{constant}, i.e. that it does \emph{not} vary with $\vv$ throughout the state transition and equals to $\sigmaw_{0}^2 = \sigmaw^2(0)$, its stable steady-state value at $\vv = 0$~\cite{EDTM2024}.

\subsection{Near-Stable-Steady-State Approximate Formulas}
\label{subsection:Near-Stable-Steady-State Approximate Formulas}


%

Former closed-form analytical formulas for the $\MTT$~\cite{Kish2002,Nobile1985} further rely on linearisation of the system dynamics ($h(\vv)$) and are thereby highly inaccurate, with discrepancies of several orders of magnitude reported for the $\MTT$ in SRAM bitcells~\cite{LASCAS2024,EDTM2024}.

Siegert~\cite{Siegert1951} derived the 
first passage time 
(FPT) 
probability of a stationary 
1D
Markov process and provided a recurrence formula for the moments of the FPT when the probability distribution satisfies a Fokker-Plank equation.
Nobile~\cite{Nobile1985} 
applied \cite{Siegert1951}'s formula to the special case of an Ornstein-Uhlenbeck process, defined by \eqref{eq:dvv/dt} with linearised drift term $h(\vv) \approx - \vv(t)/\tauo$ (full brown line in \cref{fig_MTNS2024_h}), with
\begin{equation}
\label{eq:tauo}
\frac{1}{\tauo}
= -\dv{h}{\vv}\/(0) = \dv[2]{\EE}{\vv}\/(0)
\end{equation}
the time constant of the system linearised at $\vv = 0$.
The $\MTT$ is twice the 
mean
FPT~\cite{LASCAS2024}, i.e. the mean time to reach $\vv = \Dvv$ for the first time starting from $\vv = 0$:
\begin{equation}
\label{eq:MTT Nobile}
\MTT \approx 2 \, \tauo
\,
\bigg(
\begin{aligned}[t]
&\sqrt{\pi} \int_{0}^{\ds\Dvv/(\sigmaw_{0} \sqrt{\tauo})} \diff{u} \, \exp(u^2) \\
&\mathllap{+} \int_{0}^{\ds\Dvv/(\sigmaw_{0} \sqrt{\tauo})} \diff{u} \, \exp(u^2) \erf(u)
\bigg)
\text{.}
\end{aligned}
\end{equation}

Kish's formula~\cite{Kish2002},
\begin{equation}
\label{eq:MTT Kish}
\MTT \approx \sqrt{3} \, \pi \, \tauo \, \exp \bigg( 
\frac{\Dvv^2/(2\tauo)}{\frac{1}{2} \sigmaw_{0}^2}
\bigg)
\text{,}
\end{equation}
is sometimes used in electronics~\cite{Veirano2016_journal},
and can be regarded as a simplified version of \eqref{eq:MTT Nobile}.
Within the exponential, $\Dvv^2/(2\tauo)$ is interpreted as the quasi-potential barrier evaluated at the unstable point ($\vv = \Dvv$) under the coarse near-stable-steady-state harmonic (parabolic) 
approximation (\cref{fig_MTNS2024_E}), equivalent to linearised drift. The barrier or the drift is strongly overestimated, and so is the $\MTT$.
Both \cref{eq:MTT Nobile,eq:MTT Kish}, relying on a near-stable-steady-state approximation, are inappropriate for describing the true nonlinear SRAM dynamics as previously concluded in \citet{LASCAS2024,EDTM2024} and here observed in \cref{fig_h_E}.

\subsection{Classical Eyring-Kramers Formula: Constant $\sigma^2$}
\label{subsection:Classical Eyring-Kramers Formula}

We have highlighted that the near-stable-steady-state approximation inherent to~\cite{Kish2002,Nobile1985} is twofold: the deterministic drift is linearised around $\vv =0$ only; the noise is assumed constant and also extracted at $\vv = 0$ only. Dynamic simulations invalidate these coarse simplifications, which Eyring-Kramers formula~\cite{Kramers1940}
partially alleviates 
by also taking into account the drift (quasi-potential) behaviour of the system near the \emph{unstable} point (dashed brown lines in \cref{fig_h_E})
However, it still assumes constant noise $\sigmaw(\vv) \equiv \sigmaw$.

In our notations, Eyring-Kramers formula
is written as~\cite{Berglund2011}
\begin{equation}
\begin{aligned}[b]
\label{eq:MTT Kramers0}
\MTT &\approx 2\pi\textcolor{h}{\sqrt{\tauo\,\tauM}} \, \exp \bigg( \frac{\EE(\Dvv) - \EE(0)}{\frac{1}{2} \sigmaw^2} \bigg) \\
&= 2\pi\textcolor{h}{\sqrt{\tauo\,\tauM}} \, \exp \bigg( - \frac{\int_{0}^{\Dvv} h(\vv') \,\diff{\vv'}}{\frac{1}{2} \sigmaw^2} \bigg)
\text{.}
\end{aligned}
\end{equation}
We explain how the new quantities introduced in \eqref{eq:MTT Kramers0} are defined and extracted.

$\EE(\Dvv) - \EE(0)$ is the \emph{true} quasi-potential barrier (\cref{fig_MTNS2024_E}) whose crossing by the random diffusion process leads to a state transition. Setting $\EE(0) = 0$ as arbitrary potential reference, $\EE(\Dvv)$ is obtained by numerically integrating $-h(\vv)$.
\begin{equation}
\label{eq:tauM}
\frac{1}{\tauM}
 = \dv{h}{\vv}\/(\Dvv) = \abs{\dv[2]{\EE}{\vv}\/(\Dvv)}
\end{equation}
is the time constant of the system linearised at the unstable point ($\vv = \Dvv$), straightforwardly computed from either $h(\vv)$ or $\EE(\vv)$. 
Hence, a new effective time constant of the diffusion process, defined as a geometrical mean $\tau \equiv \sqrt{\tauo \tauM}$, appears in \eqref{eq:MTT Kramers0} and accounts for the drift behaviour in both the near stable- and unstable-steady-state regions.

The proof of \eqref{eq:MTT Kramers0} is based on rigorous stochastic calculations and can be found for instance in~\cite{Berglund2011,Bouchet2016}.

Unfortunately, the classical formula \eqref{eq:MTT Kramers0} still assumes constant noise variance $\sigmaw^2$.
Using $\sigma = \sigmawo$, like in \cref{eq:MTT Nobile,eq:MTT Kish}, would amount to estimating the fluctuations near the stable steady state only.
\textit{A priori}, there is no physical or theoretical argument to justify favouring the unstable steady state by choosing $\sigma = \sigmawM$ instead.
These two exclusive choices will be shown to be inaccurate, when attempting to apply \eqref{eq:MTT Kramers0} to nonlinear stochastic bistable systems such as SRAM bitcells with varying substantially $\sigmaw(\vv)$ all along the state transition from $\vv = 0$ to $\vv = \Dvv$ 
(extracted numerical values are provided in the caption of \cref{fig_h_E}).

\section{Stochastic Formulation for Varying Noise Intensity}
\label{section:Stochastic Formulation for Varying Noise Intensity}

\subsection{Itô Drift-Diffusion Process}
\label{subsection:Itô Drift-Diffusion Process}

\emph{Itô's lemma} (see \cite[Theorem 1.1]{Ito1950} or \cite[Theorem 6]{Ito1951}) may be regarded as the chain rule for stochastic calculus, non trivial for the case of varying $\sigmaw(\vv)$.
His formula describes a new drift-diffusion process
after performing 
a change of variable $f(\vv)$ (of our choice), on which we will here leverage to incorporate the varying $\sigmaw(\vv)$ within Kramers framework:
{\color{h}
\begin{equation}
\label{eq:df Itô}
\diff{f}
= \bigg[ h(\vv) \dv{f}{\vv} + \frac{\sigmaw^2(\vv)}{2} \dv[2]{f}{\vv} \bigg] \diff{t} +  \sigmaw(\vv) \dv{f}{\vv} \diff{W}(t)
\text{.}
\end{equation}
}
Next, we let
\begin{equation}
\label{eq:f}
\sigmaw(\vv) \dv{f}{\vv} = 1
\end{equation}
as definition of $f(\vv)$ 
to obtain
{\color{h}
\begin{equation}
\label{eq:df/dt}
\diff{f}
= 
\underbrace{
\bigg[
h(\vv) \dv{f}{\vv}\/(\vv) + \frac{\sigmaw^2(\vv)}{2} \dv[2]{f}{\vv}\/(\vv)
\bigg]
}_{\ds \equiv \hh(f)} \diff{t}
+ \diff{W}(t)
\text{.}
\end{equation}
}

\subsection{Extended Eyring-Kramers Formula: $\sigmaw^2(\vv)$}
\label{subsection:Extended Eyring-Kramers Formula}

Within the formulation \eqref{eq:df/dt} in variable $f$, the noise variance is now unity because the variations of noise intensity with $\vv$ are absorbed in the drift term of increased complexity compared to \eqref{eq:dvv/dt}, as consequence if Itô's calculus rule.
\Cref{eq:MTT Kramers0} applies, however with the modified drift function $\hh(f)$ as appears in \eqref{eq:df/dt}:
\begin{equation}
\label{eq:MTT Itô}
\begin{aligned}[b]
\MTT 
&\approx 2\pi\textcolor{h}{\sqrt{\tauo\,\tauM}} \,
\exp \bigg( - 2 \int_{f(0)}^{f(\Dvv)} \hh (f') \,\diff{f'} \bigg) \\
&=  2\pi\textcolor{h}{\sqrt{\tauo\,\tauM}} \,
\exp \bigg( - 2 \int_{0}^{\Dvv} \hh (f(\vv')) \, \dv{f}{\vv}\/(\vv') \, \diff{\vv'} \bigg) \\
&
\begin{aligned}[t]
=
2\pi\textcolor{h}{\sqrt{\tauo\,\tauM}} \, \exp \bigg( 
- 2 & \int_{0}^{\Dvv} 
\bigg[
h(\vv') \dv{f}{\vv}\/(\vv') \\
&
+ \frac{\sigmaw^2(\vv')}{2} \dv[2]{f}{\vv}\/(\vv')
\bigg]
\bigg)
\dv{f}{\vv}\/(\vv') \, \diff{\vv'}
\text{.}
\end{aligned}
\end{aligned}
\end{equation}
From \eqref{eq:f}, we easily compute
{\color{h}
\begin{equation}
\label{eq:df/dvv d2f/dvv2}
\dv{f}{\vv}\/(\vv) = \frac{1}{\sigmaw(\vv)} 
\quad \text{ and } \quad
\dv[2]{f}{\vv}\/(\vv) = -\frac{\ds\dv{\sigmaw}{\vv}\/(\vv)}{\sigmaw^2(\vv)}
\end{equation}
}
and \eqref{eq:MTT Itô} becomes, after splitting the two terms within the exponential
\begin{equation*}
\MTT 
\begin{aligned}[t]
\approx 2\pi\textcolor{h}{\sqrt{\tauo\,\tauM}} 
&\, \exp 
\bigg( 
- 2 \int_{0}^{\Dvv}
\frac{h(\vv')}{\sigmaw^2(\vv')}
\, \diff{\vv'}
\bigg) \\
&\cdot \exp 
\bigg( 
\underbrace{
- 2 \int_{0}^{\Dvv}
\frac{1}{2} \frac{\dv{\sigmaw}{\vv}\/(\vv')}{\sigmaw(\vv')}
\, \diff{\vv'}
}_{\ds - \ln\Big(\frac{\sigmaw(\Dvv)}{\sigmaw(0)}\Big)}
\bigg)
\text{.}
\end{aligned}
\end{equation*}
The general Eyring-Kramers formula, now accounting for both nonlinear drift and state-dependent white noise variance, finally reads:
\begin{equation}
\label{eq:MTT Kramers sigma(v)}
\MTT \approx 2\pi\textcolor{h}{\sqrt{\tauo\,\tauM}} 
\, \frac{\sigmawo}{\sigmawM}
\, \exp 
\bigg( 
- 2 \int_{0}^{\Dvv}
\frac{h(\vv')}{\sigmaw^2(\vv')}
\, \diff{\vv'}
\bigg)
\text{,}
\end{equation}
with $\sigmawo \equiv \sigmaw(0)$ and $\sigmawM \equiv \sigmaw(\Dvv)$.

\subsection{Approximate Extended Eyring-Kramers Formula: 
Near-Steady-States 
$\sigmawo$ and $\sigmawM$}
\label{subsection:Approximate Extended Eyring-Kramers Formula}

The noise can sometimes only be caracterised through the random time fluctuations of observable states of the system, and hence the variance extracted at the steady states only. That is, one cannot generally assume that the noise intensity $\sigmaw(\vv)$ as appears in \eqref{eq:MTT Kramers sigma(v)} is available for all $\vv$, only that $\sigmawo$ and $\sigmawM$ can be inferred from the steady-state fluctuations of $\vv$.
The link between the intrinsic noise ($\sigmaw(\vv) w(t)$) and the state ($\vv(t)$) is, at equilibrium, given by the fluctuation–dissipation theorem~\cite{Kubo1966}.


To address this situation where the noise source is partially hidden, we also propose an approximate version of \eqref{eq:MTT Kramers sigma(v)}.
Considering that only $\sigmawo^2$ and $\sigmawM^2$ are known, we assume a linear interpolation of $1/\sigmaw^2(\vv)$ between $\vv = 0$ and $\vv = \Dvv$:
\begin{equation}
\label{eq:sigma(vv)}
\frac{1}{\sigmaw(\vv)^2}
\approx
\frac{1}{\sigmawo^2}
+
\bigg(
\frac{1}{\sigmawM^2} - \frac{1}{\sigmawo^2}
\bigg)
\frac{\vv}{\Dvv}
\text{.}
\end{equation}
This choice, as well as being natural and very convenient from an analytical perspective given the integral in \eqref{eq:MTT Kramers sigma(v)}, is physically sound since the intensity of physical noise sources evolves smoothly and gradually with the state. 
{\color{h} Should we have, beyond $\sigmawo^2$ and $\sigmawM^2$, some additional information about $1/\sigmaw^2(\vv)$, notably its convexity, a parabolic approximation could for instance be more accurate.}
Substituting \eqref{eq:sigma(vv)} into \eqref{eq:MTT Kramers sigma(v)} leads, after integration by parts (Appendix~\ref{Appendix}), to our main result:
\begin{equation}
\label{eq:MTT Kramers sigma(v) interp}
\MTT \approx 2\pi\textcolor{h}{\sqrt{\tauo\,\tauM}}
\, \frac{\sigmawo}{\sigmawM}
\, \exp \bigg( \frac{\EE(\Dvv)}{\frac{1}{2} \sigmawM^2} \bigg)
\, \exp \bigg( - 2
\bigg( \frac{1}{\sigmawM^2} - \frac{1}{\sigmawo^2} \bigg)
\EEav \bigg)
\text{,}
\end{equation}
where
\begin{equation}
\label{eq:EEav}
\EEav
\equiv
\frac{1}{\Dvv} \int_{0}^{\Dvv} \EE(\vv') \,\diff{\vv'}
\text{.}
\end{equation}
We refer to \eqref{eq:MTT Kramers sigma(v) interp} as the \emph{approximate extended Eyring-Kramers formula} since it does incorporate the noise behaviour near both the stable and the unstable steady states, unlike the classical formula \eqref{eq:MTT Kramers0} restricted to a single constant $\sigma$.

We 
interpret \eqref{eq:MTT Kramers sigma(v) interp}.
Beyond the time constant prefactor $2\pi/\sqrt{\tauo\,\tauM}$ as present in the classical Eyring-Kramers formula, three factors appear:
\begin{itemize}
\item
$\sigmawo/\sigmawM$ results from the additional drift term in Itô's formula.
It is expected to have a negligible impact on the order of magnitude of the $\MTT$, compared to the exponentials that follow.
\item The factor
\[
\exp \bigg( \frac{\EE(\Dvv)}{\frac{1}{2} \sigmawM^2} \bigg)
\]
is nothing but Eyring-Kramers exponential factor, however with $\sigma = \sigmawM$ used instead of $\sigmawo$.
\item The factor
\[
\exp \bigg( - 2
\bigg( \frac{1}{\sigmawM^2} - \frac{1}{\sigmawo^2} \bigg)
\EEav \bigg)
\]
may be regarded as a factor correcting for the fact that the noise variance is not constant throughout the state transition.
It is trivially unity when $\sigmaw(\vv) = \sigmawo = \sigmawM = \text{constant}$.
If $\sigmawM < \sigmawo$, the argument within the exponential is \emph{positive} and the factor is larger than $1$, thereby compensating the excessive first exponential involving a too small $\sigmaw$.
Conversely, if $\sigmawM > \sigmawo$, the correcting factor is smaller than $1$.
\end{itemize}

As will be evidenced in \cref{section:Predictions and Discussion}, we harness this extended law \eqref{eq:MTT Kramers sigma(v) interp} to predict the $\MTT$ in SRAM bitcells with much greater accuracy than the former analytical formulas reviewed in \cref{section:Previous Works}. 
Whenever $\sigmaw(\vv)^2$ is explicitly available or characterisable for other stochastic bistable systems, the authors naturally encourage the use of the most general formula \eqref{eq:MTT Kramers sigma(v)}.

\section{Predictions and Discussion}
\label{section:Predictions and Discussion}

\begin{figure}[]
\newcommand\myfontsize{\small}
\newcommand\mytickfontsize{\footnotesize}
\myfontsize
\psfragscanon
\psfrag{VDD [mV]}[cc][cc]{\myfontsize$\VDD \, [\si{\milli\volt}]$}
\psfrag{MTTF [s]}[cc][cc]{\myfontsize$MTT \, [\si{\second}]$}
\psfrag{60}[cc][cc]{\mytickfontsize$60$}
\psfrag{70}[cc][cc]{\mytickfontsize$70$}
\psfrag{80}[cc][cc]{\mytickfontsize$80$}
\psfrag{e6}[cr][cr]{\mytickfontsize$10^{6}$}
\psfrag{e3}[cr][cr]{\mytickfontsize$10^{3}$}
\psfrag{e0}[cr][cr]{\mytickfontsize$1$}
\psfrag{em1}[cr][cr]{\mytickfontsize$10^{-1}$}
\psfrag{em2}[cr][cr]{\mytickfontsize$10^{-2}$}
\psfrag{em3}[cr][cr]{\mytickfontsize$10^{-3}$}
\psfrag{em4}[cr][cr]{\mytickfontsize$10^{-4}$}
\psfrag{em5}[cr][cr]{\mytickfontsize$10^{-5}$}
\psfrag{em6}[cr][cr]{\mytickfontsize$10^{-6}$}
\psfrag{em7}[cr][cr]{\mytickfontsize$10^{-7}$}
\psfrag{em8}[cr][cr]{\mytickfontsize$10^{-8}$}
\psfrag{em9}[cr][cr]{\mytickfontsize$10^{-9}$}
\psfrag{NOISETRAN}[cl][cl]{\color{b}\textbf{Dynamical simulations}}
\psfrag{Kish}[cl][tl]{\color{r}Kish \eqref{eq:MTT Kish}~\cite{Kish2002}}
\psfrag{Nobile}[cl][tl]{\color{orange}Nobile \eqref{eq:MTT Nobile}~\cite{Nobile1985}}
\psfrag{Kramers M}[cl][cl]{\color{chocolateBrown} Kramers \eqref{eq:MTT Kramers0} with $\sigmawM$}
\psfrag{Kramers interp}[cl][cl]{\color{violet} \textbf{Extended Kramers} \eqref{eq:MTT Kramers sigma(v) interp}}
\psfrag{Kramers 0}[cl][cl]{\color{mybrown} Kramers \eqref{eq:MTT Kramers0} with $\sigmawo$}
\vspace{3mm}
\includegraphics[scale=1]{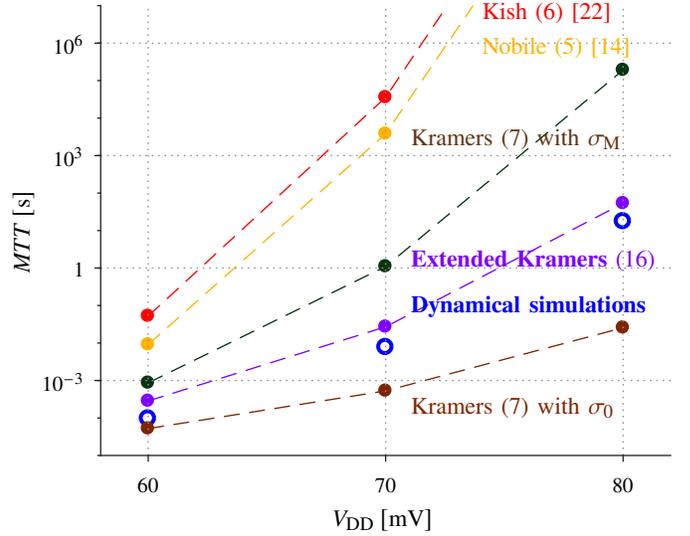}
\caption{
Comparison between the $\MTT$ estimated empirically from expensive transient noise simulations~\cite{SSE2023}, 
and from predictions of analytical near-equilibrium formulas and of the extended Eyring-Kramers formula.
}
\label{fig_MTT}
\end{figure}

In this section, we assess the respective accuracies of the different closed-form formulas for the $\MTT$ (\cref{fig_MTT}) and point out the origins of the discrepancies. 
We consider three different SRAM bitcells, distinguished by different values of the supply voltage ($\VDD$) parameter whose dominant effect is to reduce the distance $\Dvv$ between stable steady state and unstable state and hence to lower the $\MTT$.
References values for the $\MTT$ are provided by massive and accurate dynamic Monte-Carlo simulations from~\cite{SSE2023}, based on a multidimensional model similar in mind to \eqref{eq:2D}.

As reviewed in \cref{subsection:Near-Stable-Steady-State Approximate Formulas}, \cite{Kish2002,Nobile1985} assume both linearised drift and constant $\sigmaw_{0}^2$ related to the stable steady-state $\tauo$ only) and are totally inadequate to describe the SRAM dynamics with strongly nonlinear and non-monotonic drift function $h(\vv)$ (\cref{fig_MTNS2024_h}) and intrinsic noise variance which depends on the time-evolving state $\vv$ for nonlinear electronic devices.
The discrepancy can reach orders of magnitude, as seen in \cref{fig_MTT} and already reported for another case study in~\cite{LASCAS2024,EDTM2024}.

The classical Eyring-Kramers formula (\cref{subsection:Classical Eyring-Kramers Formula}) is rewarded for its nonlinear treatment of the drift $h(\vv)$ 
by slightly
improved
accuracy.
However, the noise is either under- or over-estimated, respectively when $\sigmawM$ or $\sigmawo$ is used in \eqref{eq:MTT Kramers0}, causing the $\MTT$ to be either over- or under-estimated compared to Monte-Carlo.

Our extended Eyring-Kramers formula 
was constructed in \cref{subsection:Extended Eyring-Kramers Formula} to remedy these important shortcomings.
It achieves, at least for the three investigated cases, a remarkable accuracy when compared to Monte-Carlo. Importantly, the trend $\MTT$ versus $\VDD$ seems properly captured by our model, as a result of also taking into account the drift-diffusion behaviours near the unstable point.
The residual discrepancy noticed in \cref{fig_MTT} is likely attributed to the linear interpolation of $1/\sigmaw^2(\vv)$, which could implicitly inflate or reduce the actual noise intensity depending on whether the true hidden $\sigmaw^2(\vv)$ is a concave or convex function.
{\color{h}
We may indeed observe that, for the case illustrated through this paper, \eqref{eq:MTT Kramers sigma(v) interp} systematically overestimates the $\MTT$, thereby suggesting that the true $1/\sigmaw^2(\vv)$ is some 
nonlinear
function slightly overestimated by the linear interpolation.
}

\section{Conclusions and Perspectives}
\label{section:Conclusions and Perspectives}

Efficient and accurate prediction methodology of the mean transition time in bistable autonomous systems finds applications in various fields. Here, we have focused on SRAM bitcells as integrated in electronic chips for data retention.
The stochastic nonlinear dynamical model may be thought as formally describing the overdamped motion of a Brownian subjected to drift (deriving from a potential) and diffusion (noise intrinsic to dissipative devices).
Hence, apart from the notations, the extended Eyring-Kramers formulas for the $\MTT$ proposed in this paper, either the general \eqref{eq:MTT Kramers sigma(v)} or the approximate \eqref{eq:MTT Kramers sigma(v) interp}, can be straightforwardly applied to a wide class of drift-diffusion problems as referenced in the introduction.
Whereas \eqref{eq:MTT Kramers sigma(v)} address the most general case of arbitrary non-constant white noise variance, \eqref{eq:MTT Kramers sigma(v) interp} is valuable when the stochasticity of the system is partially hidden and 
can only be identified
near the stable steady state and the unstable state.
Our first numerical trials give faith in the efficiency and accuracy of this approach, which is still awaiting further validation across more case studies.



\appendices

\section{Derivation of \eqref{eq:MTT Kramers sigma(v) interp}}
\label{Appendix}

We expand \eqref{eq:MTT Kramers sigma(v)} 
assuming that
$1/\sigmaw^2(\vv)$ is linearly interpolated \textcolor{h}{between} $\vv = 0$ and $\vv = \Dvv$ 
according to 
\eqref{eq:sigma(vv)}:
\begin{equation}
\label{eq:interp}
\begin{aligned}[b]
\MTT 
&
\begin{aligned}[t]
\approx 2\pi\textcolor{h}{\sqrt{\tauo\,\tauM}} 
\, \frac{\sigmawo}{\sigmawM}
\, \exp 
&
\bigg( 
- 2
\bigg[
\int_{0}^{\Dvv}
h(\vv')
\,
\frac{1}{\sigmawo^2}
\, \diff{\vv'} \\
&+ \int_{0}^{\Dvv}
h(\vv')
\,
\bigg(
\frac{1}{\sigmawM^2} - \frac{1}{\sigmawo^2}
\bigg)
\frac{\vv}{\Dvv}
\, \diff{\vv'}
\bigg]
\bigg)
\text{.}
\end{aligned} \\
\end{aligned}
\end{equation}
The first integral within the exponential is 
exactly the same as
\eqref{eq:MTT Kramers0} with $\sigma = \sigmawo$ and $\EE(0) \equiv 0$ by convention:
\begin{equation}
\label{eq:int1}
\exp \bigg( 
- 2 \int_{0}^{\Dvv}
h(\vv')
\,
\frac{1}{\sigmawo^2}
\, \diff{\vv'}
\bigg)
=
\exp \bigg( \frac{\EE(\Dvv)}{\frac{1}{2} \sigmawo^2} \bigg)
\end{equation}
We integrate the second term by parts, remembering that $h(\vv) = -\diff{\EE}/\diff{\vv}$:
\begin{equation}
\label{eq:int2}
\begin{aligned}
\int_{0}^{\Dvv} h(\vv') \, \vv \, \diff{\vv'}
&= \big[ - \EE(\vv)\, \vv \big]_{0}^{\Dvv}
- \int_{0}^{\Dvv} - \EE(\vv') \, \diff{\vv'} \\
&= - \EE(\Dvv) \Dvv + \int_{0}^{\Dvv} \EE(\vv') \, \diff{\vv'} \\
&=
- \big( \EE(\Dvv) - \EEav \big) \Dvv
\end{aligned}
\end{equation}
where the average $\EEav$ is defined in \eqref{eq:EEav}.

Inserting \cref{eq:int1,eq:int2} in \eqref{eq:interp} yields
\begin{equation}
\label{eq:interp 2}
\begin{aligned}[b]
\MTT 
&
\begin{aligned}[t]
\approx 2\pi\textcolor{h}{\sqrt{\tauo\,\tauM}} 
\, \frac{\sigmawo}{\sigmawM}
& \, 
\exp \bigg( \frac{\EE(\Dvv)}{\frac{1}{2} \sigmawo^2} \bigg) \\
& \cdot
\exp 
\bigg( 
+ 2
\bigg(
\frac{1}{\sigmawM^2} - \frac{1}{\sigmawo^2}
\bigg)
\big( \EE(\Dvv) - \EEav \big)
\bigg)
\text{.}
\end{aligned} \\
&
\begin{aligned}[t]
= 2\pi\textcolor{h}{\sqrt{\tauo\,\tauM}} 
\, \frac{\sigmawo}{\sigmawM}
& \, 
\exp \bigg( \frac{\EE(\Dvv)}{\frac{1}{2} \sigmawM^2} \bigg) \\
& \cdot
\exp \bigg( - 2
\bigg( \frac{1}{\sigmawM^2} - \frac{1}{\sigmawo^2} \bigg)
\EEav \bigg)
\bigg)
\text{.}
\end{aligned}
\end{aligned}
\end{equation}
which is \eqref{eq:MTT Kramers sigma(v) interp}.


\bibliographystyle{IEEEtran}
\bibliography{IEEEabrv,bib}

\begin{thebibliography}{10}
\providecommand{\url}[1]{#1}
\csname url@samestyle\endcsname
\providecommand{\newblock}{\relax}
\providecommand{\bibinfo}[2]{#2}
\providecommand{\BIBentrySTDinterwordspacing}{\spaceskip=0pt\relax}
\providecommand{\BIBentryALTinterwordstretchfactor}{4}
\providecommand{\BIBentryALTinterwordspacing}{\spaceskip=\fontdimen2\font plus
\BIBentryALTinterwordstretchfactor\fontdimen3\font minus
  \fontdimen4\font\relax}
\providecommand{\BIBforeignlanguage}[2]{{%
\expandafter\ifx\csname l@#1\endcsname\relax
\typeout{** WARNING: IEEEtran.bst: No hyphenation pattern has been}%
\typeout{** loaded for the language `#1'. Using the pattern for}%
\typeout{** the default language instead.}%
\else
\language=\csname l@#1\endcsname
\fi
#2}}
\providecommand{\BIBdecl}{\relax}
\BIBdecl

\bibitem{SSE2023}
L.~Van~Brandt, F.~Silveira, J.-C. Delvenne, and D.~Flandre, ``{On Noise-Induced
  Transient Bit Flips in Subthreshold SRAM},'' \emph{Solid-State Electronics},
  vol. 208, p. 108715, 2023.

\bibitem{Dhilnikov2005}
A.~Shilnikov, R.~L. Calabrese, and G.~Cymbalyuk, ``Mechanism of bistability:
  tonic spiking and bursting in a neuron model,'' \emph{Physical Review E},
  vol.~71, no.~5, p. 056214, 2005.

\bibitem{Goldbeter2018}
A.~Goldbeter, ``Dissipative structures in biological systems: bistability,
  oscillations, spatial patterns and waves,'' \emph{Philosophical Transactions
  of the Royal Society A: Mathematical, Physical and Engineering Sciences},
  vol. 376, no. 2124, p. 20170376, 2018.

\bibitem{Livina2010}
V.~N. Livina, F.~Kwasniok, and T.~M. Lenton, ``Potential analysis reveals
  changing number of climate states during the last 60 kyr,'' \emph{Climate of
  the Past}, vol.~6, no.~1, pp. 77--82, 2010.

\bibitem{Hirota2011}
M.~Hirota, M.~Holmgren, E.~H. Van~Nes, and M.~Scheffer, ``Global resilience of
  tropical forest and savanna to critical transitions,'' \emph{Science}, vol.
  334, no. 6053, pp. 232--235, 2011.

\bibitem{Vellela2009}
M.~Vellela and H.~Qian, ``Stochastic dynamics and non-equilibrium
  thermodynamics of a bistable chemical system: the {Schl{\"o}gl} model
  revisited,'' \emph{Journal of The Royal Society Interface}, vol.~6, no.~39,
  pp. 925--940, 2009.

\bibitem{LASCAS2024}
L.~Van~Brandt, J.-C. Delvenne, and D.~Flandre, ``{Variability-Aware
  Noise-Induced Dynamic Instability of Ultra-Low-Voltage SRAM Bitcells},'' in
  \emph{IEEE LASCAS 2024, Punta del Este, Uruguay}, 2024.

\bibitem{EDTM2024}
L.~Van~Brandt, D.~Flandre, and J.-C. Delvenne, ``{Stochastic Nonlinear
  Dynamical Modelling of SRAM Bitcells in Retention Mode},'' in \emph{IEEE EDTM
  2024, Bangalore, India}, 2024, invited talk.

\bibitem{Berglund2011}
N.~Berglund, ``Kramers' law: Validity, derivations and generalisations,''
  \emph{arXiv preprint arXiv:1106.5799}, 2011.

\bibitem{Freitas2022_reliability}
N.~Freitas, K.~Proesmans, and M.~Esposito, ``Reliability and entropy production
  in nonequilibrium electronic memories,'' \emph{Physical Review E}, vol. 105,
  no.~3, p. 034107, 2022.

\bibitem{Rezaei2020}
E.~Rezaei, M.~Donato, W.~R. Patterson, A.~Zaslavsky, and R.~I. Bahar,
  ``{Fundamental Thermal Limits on Data Retention in Low-Voltage CMOS Latches
  and SRAM},'' \emph{IEEE Transactions on Device and Materials Reliability},
  vol.~20, no.~3, pp. 488--497, 2020.

\bibitem{Weidenmuller1984}
H.~Weidenm{\"u}ller and Z.~Jing-Shang, ``Stationary diffusion over a
  multidimensional potential barrier: A generalization of {Kramers}' formula,''
  \emph{Journal of statistical physics}, vol.~34, pp. 191--201, 1984.

\bibitem{Siegert1951}
A.~J. Siegert, ``On the first passage time probability problem,''
  \emph{Physical Review}, vol.~81, no.~4, p. 617, 1951.

\bibitem{Nobile1985}
A.~Nobile, L.~Ricciardi, and L.~Sacerdote, ``{Exponential trends of
  Ornstein--Uhlenbeck first-passage-time densities},'' \emph{Journal of Applied
  Probability}, vol.~22, no.~2, pp. 360--369, 1985.

\bibitem{Kramers1940}
H.~A. Kramers, ``Brownian motion in a field of force and the diffusion model of
  chemical reactions,'' \emph{Physica}, vol.~7, no.~4, pp. 284--304, 1940.

\bibitem{Bouchet2016}
F.~Bouchet and J.~Reygner, ``Generalisation of the eyring--kramers transition
  rate formula to irreversible diffusion processes,'' in \emph{Annales Henri
  Poincar{\'e}}, vol.~17, no.~12.\hskip 1em plus 0.5em minus 0.4em\relax
  Springer, 2016, pp. 3499--3532.

\bibitem{Johnson1928}
J.~B. Johnson, ``Thermal agitation of electricity in conductors,''
  \emph{Physical review}, vol.~32, no.~1, p.~97, 1928.

\bibitem{Nyquist1928}
H.~Nyquist, ``Thermal agitation of electric charge in conductors,''
  \emph{Physical review}, vol.~32, no.~1, p. 110, 1928.

\bibitem{Kubo1966}
R.~Kubo, ``The fluctuation-dissipation theorem,'' \emph{Reports on progress in
  physics}, vol.~29, no.~1, p. 255, 1966.

\bibitem{APL2023}
L.~Van~Brandt and J.-C. Delvenne, ``Noise-dissipation relation for nonlinear
  electronic circuits,'' \emph{Applied Physics Letters}, vol. 122, no.~26,
  2023.

\bibitem{Zhang2006}
B.~Zhang, A.~Arapostathis, S.~Nassif, and M.~Orshansky, ``{Analytical Modeling
  of SRAM Dynamic Stability},'' in \emph{Proceedings of the 2006 IEEE/ACM
  international conference on Computer-aided design}, 2006, pp. 315--322.

\bibitem{Kish2002}
L.~B. Kish, ``{End of Moore's law: thermal (noise) death of integration in
  micro and nano electronics},'' \emph{Physics Letters A}, vol. 305, no. 3-4,
  pp. 144--149, 2002.

\bibitem{Rogal2021}
J.~Rogal, ``Reaction coordinates in complex systems-a perspective,'' \emph{The
  European Physical Journal B}, vol.~94, pp. 1--9, 2021.

\bibitem{Veirano2016_journal}
F.~Veirano, F.~Silveira, and L.~Naviner, ``{Minimum Operating Voltage Due to
  Intrinsic Noise in Subthreshold Digital Logic in Nanoscale CMOS},'' \emph{J.
  of Low Power Electronics}, vol.~12, no.~1, pp. 74--81, 2016.

\bibitem{Ito1950}
K.~It{\^o}, ``Stochastic differential equations in a differentiable manifold,''
  \emph{Nagoya Mathematical Journal}, vol.~1, pp. 35--47, 1950.

\bibitem{Ito1951}
------, ``On a formula concerning stochastic differentials,'' \emph{Nagoya
  Mathematical Journal}, vol.~3, pp. 55--65, 1951.

\end{thebibliography}

\end{document}